# • CAPÍTULO 24 •

## Evaluación de una antena con tecnología de guía de onda integrada al substrato para redes vehículares

Kevin Delgadillo,
César Rodríguez,
Wilder Castellanos,
Héctor Guarnizo

## 1. Introducción

Actualmente las nuevas arquitecturas de redes inalámbricas están generando el surgimiento de nuevos escenarios de aplicación, tal es el caso de las redes móviles ad hoc (también conocidas como MANETs, Mobile Ad doc NETworks)[1]. Las MANETs se caracterizan por que sus nodos pueden establecer entre ellos comunicación directa, sin necesidad de nodos centrales que cumplan las funciones de administración y control. Este hecho permite que las redes MANETs se puedan desplegar sin infraestructura fija, lo cual supone un ahorro en costos de implementación. Sin embargo, es importante anotar que los nodos deben autogestionar el acceso, el enrutamiento y el mantenimiento de la conectividad, lo cual conlleva a que los nodos, no solo sirvan de equipos terminales, sino que además deben cumplir con tareas de enrutadores. Para cumplir con estas tareas, en los últimos años se han desarrollado novedosos protocolos y tecnologías para las redes MANETs [2]–[4], que proyectan dichas redes como la plataforma base sobre la cual se implementan escenarios como IoT (Internet of Things) [5], las ciudades inteligentes (Smart cities) [6], [7] e IoP (Internet of People) [8]. Un tipo particular de las redes MANET son las VANETs (Vehicular Ad-hoc Networks) o redes vehiculares. Las VANETs plantean una infraestructura tecnológica que posibilitan el desarrollo práctico de las ciudades inteligentes [9], [10] ya que hacen posible que los vehículos actúen como los nodos de una red para la transmisión de datos, voz y video. Dicha comunicación puede establecerse desde vehículos a vehículos (conocida como comunicación V2V) o entre vehículos e infraestructura (también conocida como V2I). El principal propósito de este tipo de redes es mejorar la seguridad del pasajero, así como intercambiar información sobre el tráfico y los puntos de servicio cercanos, tales como: estaciones de gasolina, hospitales, restaurantes, entre otros.
Las VANETs también emergen como una herramienta tecnológica útil para organizar la caótica movilidad de las ciudades modernas. Por intermedio de estas redes, no solo los conductores pueden estar permanentemente informados sobre el estado de las vías, sino que además las entidades encargadas de la gestión de la movilidad pueden controlar en tiempo real el tráfico de la ciudad. Sin embargo, algunas limitaciones deben ser superadas para que la implementación de las VANETs sea una realidad. En primer lugar, es necesario el desarrollo de nuevos protocolos de comunicación que incluyan mecanismos que permitan establecer y mantener los enlaces inalámbricos a pesar de la alta movilidad de los nodos. Por otra parte, es necesario la implementación del hardware necesario para soportar dichas comunicaciones sin sacrificar el espacio dentro del vehículo. Teniendo en cuenta este aspecto, la tecnología SIW (Substrate Integrated Waveguide, guía de onda integrada al substrato) [11], [12] surge como una de las soluciones en la construcción de estos nodos inalámbricos vehiculares. Esta tecnología permite la construcción de antenas simples, de tamaño pequeño y a bajo costo, lo cual las hace apropiadas para las VANETs.

En este artículo se presenta el diseño y la evaluación de una antena construida con tecnología SIW operando a una frecuencia de 2.4 GHz. El diseño de la antena fue realizado mediante el software ANSYS HFSS, por medio del cual también se evaluó el patrón de radiación, la frecuencia de operación y el ancho de banda. A partir de los resultados obtenidos con la antena, se realizó una simulación de una VANET sobre la plataforma NS2 (Network Simulator 2). Los nodos de las VANET fueron configurados con los parámetros de la antena SIW desarrollada. Dicho estudio basado en simulaciones tuvo como propósito, evaluar la integración de este tipo de antenas en los nodos vehiculares, evaluando algunas métricas como el flujo de datos y la tasa de paquetes perdidos. Con el fin de hacer las simulaciones de las VANETs más realistas, fueron utilizadas algunas herramientas complementarias a NS2, como MOVE (Mobility model generator for Vehicular networks) [13] y SUMO (Simulation of Urban Mobility) [14], las cuales fueron utilizadas para generar las rutas y el movimiento de los vehículos, respectivamente. Los resultados obtenidos en las simulaciones muestran las difíciles condiciones de comunicación dentro de las VANETs debido principalmente a la movilidad de los nodos. Esto produce pérdidas de conectividad temporales que hacen que aumenten las pérdidas de paquetes y el retardo en la transmisión. Sin embargo, el uso de las antenas con tecnología SIW sigue siendo muy promisorio debido a la buena ganancia que presentan estas antenas, lo cual favorece el establecimiento de los enlaces inalámbricos entre los nodos.

## 2. Estado del arte

El desarrollo de las VANETs ha sido un área de investigación muy dinámica en los últimos años. El trabajo ha estado principalmente enfocado al desarrollo de nuevos protocolos de comunicación y a la implementación de mecanismos de calidad de servicio que contribuyan a la transmisión efectiva de tráfico multimedia. Estos trabajos investigativos han sido validados, principalmente, a través de escenarios simulados. Por lo tanto, la simulación de una red VANET se ha convertido en un importante tema de investigación. Por ejemplo, en la referencia [10] se analiza el desarrollo de un nuevo simulador de estas redes, y advierten que para el desarrollo de este tipo de simulaciones es necesario contar con un modelo de movilidad de un vehículo, y un simulador de redes. De acuerdo a los autores de dicho trabajo, aunque existan simuladores independientes para estos dos aspectos, el lograr juntarlos es algo muy complejo. Uno de los estudios de simulación de redes VANETs más relevante es el proyecto TraNS (Traffic and Network Simulator Environment) [15], el cual es un entorno de simulación basado en el software de simulación NS2 (Network Simulator 2) y el simulador de movilidad SUMO. De igual formar varios trabajos de investigación han estado centrados en la generación de modelos de movilidad destinados a la generación de escenarios más realistas, para desarrollar sobre ellos los estudios de simulación [16], [17].

Por otro lado, el desarrollo de nuevos mecanismos que permitan la transmisión de flujos más sensibles a las variaciones de la red, también ha sido objeto de numerosos estudios. Por ejemplo, el estudio presentado en [18] aborda el problema de la transmisión de video sobre una red VANET mediante el análisis de diferentes protocolos de enrutamiento. Esto con el fin de evaluar la eficiencia de cada uno de los protocolos estudiados. Los protocolos analizados en dicho trabajo fueron: AODV (Ad hoc On-demand Distance Vector), OLSR (Optimized Link State Routing) y CLWPR (Cross-Layer Optimized Position based Routing). OLSR y CLWPR fueron los protocolos que mostraron mejor cobertura debido a que permitieron una efectiva comunicación a más 600m, mientras que con el protocolo AODV solo se tuvo un alcance efectivo a los 400m. En cuanto al retardo, también los protocolos OLSR y CLWPR obtuvieron mejores resultados (un retardo de 7ms independiente de la densidad de autos y tasa de transmisión) frente a un retardo promedio de 10ms alcanzado con AODV y con densidad media de vehículos.

Por otra parte, las antenas integradas al sustrato llevan varios años de desarrollo. Algunos ejemplos de antenas implementadas bajo la técnica de guía de onda integrada al sustrato (SIW), se pueden consultar en las referencias [19], [20]. Estos estudios presentan el diseño de antenas planares que tiene una cavidad y un elemento de alimentación que son completamente construidos sobre un sustrato simple. Además presentan las ventajas de la antena planar convencional como su bajo peso, pequeñas dimensiones, fabricación de bajo costo y la integración con un circuito planar [21]. Las antenas ranuradas pueden reducir el tamaño gracias a las condiciones de frontera y así mismo se reduce el tamaño de la antena resonante. Para conseguir dicha condición de frontera los dos cortocircuitos en el extremo de la ranura resonante se remplazan por un reactivo de frontera, que incluye cargas capacitivas e inductivas. En la referencia [22] se muestra el procedimiento para diseñar una antena ranurada para cualquier tamaño. Dicho procedimiento se basa en un circuito equivalente para la antena y la estructura de alimentación, los parámetros del circuito se extraen usando un modelo directo de onda completa junto con un algoritmo genético. Estos parámetros se emplearon para encontrar una coincidencia perfecta con una línea de 50 ohmios y la antena ranurada tuvo unas dimensiones de $0.05\lambda_0$ X $0.05\lambda_0$, obteniéndose una ganancia de -3dBi con solamente pérdidas óhmicas.

Una de las principales ventajas de las antenas con tecnología SIW es su tamaño reducido. Esta característica las hace ideales para la fabricación de sensores inalámbricos, tal como lo muestra el trabajo presentado en [23]. En dicho trabajo se plantea el diseño de una antena para un sensor RFID (Radio Frequency IDentification). La geometría de dicha antena se basa en un perfil de ranura meandro en un parche suspendido, esto permite el alojamiento de sensores y circuitos electrónicos en espacios más pequeños. En este diseño se empleó un algoritmo genético para optimizar los parámetros geométricos, además de maximizar la ganancia de la antena.

## 3. Las Redes Vehiculares (VANET)

Las VANETs se consideran como una especialización de las redes móviles ad-hoc (MANET). La principal diferencia es que en MANET los nodos pueden moverse a lo largo y ancho del escenario mientras que en las redes VANET cada vehículo se define como un nodo de red equipado con una unidad de comunicación denominada OBU (on- Board Unit) y una de aplicación llamada AU (Application Unit) [24]. La principal función de la OBU es el intercambio de datos con otros vehículos o con los puntos de acceso estacionarios que se encuentran sobre la carretera, llamados RSU (Road-Side Unit).

En una VANET, todos los elementos que conforman la red crean dominios entre ellos. Esto permite el establecimiento de la comunicación entre los diferentes nodos de una forma más organizada. Los diferentes dominios pueden clasificarse dependiendo del funcionamiento, en [24]:

- **Dominio de vehículo:** se conforma por la OBU y las AU del nodo, forman una red bidireccional dentro del nodo (vehículo) y permite la conexión inalámbrica o cableada de estos.
- **Dominio Ad-hoc:** es la comunicación inalámbrica creada para conectar dos nodos o más entre sí o los nodos con la RSU. Dicha comunicación se realiza bajo el estándar IEEE 802.11p de la IEEE, con algunas variaciones en los algoritmos de seguridad, networking, gestión de recursos y multicanalidad, entre otras modificaciones. También es posible el uso de otras tecnologías inalámbricas como WIFI, 3G, LTE, entre otros.
- **Dominio de infraestructura:** este se forma mediante las redes de acceso y la infraestructura que soporta el acceso a internet, solicitado por los nodos o las RSU.

Los dominios que se explicaron anteriormente se pueden observar en la Figura 1, la cual muestra los componentes de la red VANET. Las comunicaciones que se establecen en las VANETS pueden ser V2V, que es la comunicación inter vehicular o vehículo a vehículo, donde los automóviles intercambian mensajes directamente. También se establece la comunicación entre los vehículos y la infraestructura, llamada V2I. Este tipo de comunicación es la establecida con dispositivos fijos como los peajes y los puntos de acceso a internet [25].

Las VANETs poseen una topología altamente dinámica, esto quiere decir, que para una red VANET es complejo definir una topología específica debido a que depende específicamente de los vehículos que la conforman. Además, los nodos se encuentran en movimiento continuo y la comunicación ya sea V2V o V2I se realiza en un tiempo muy corto, lo cual dificulta la identificación de la topología [25]. Otra característica es que poseen canales de comunicación variables en tiempo y frecuencia. Esto se produce debido a los obstáculos (como árboles, edificios, muros y demás) que van apareciendo durante el recorrido de los vehículos. Estos obstáculos ocasionan pérdidas de la señal en tiempo y frecuencia, con mayor intensidad que las redes móviles.

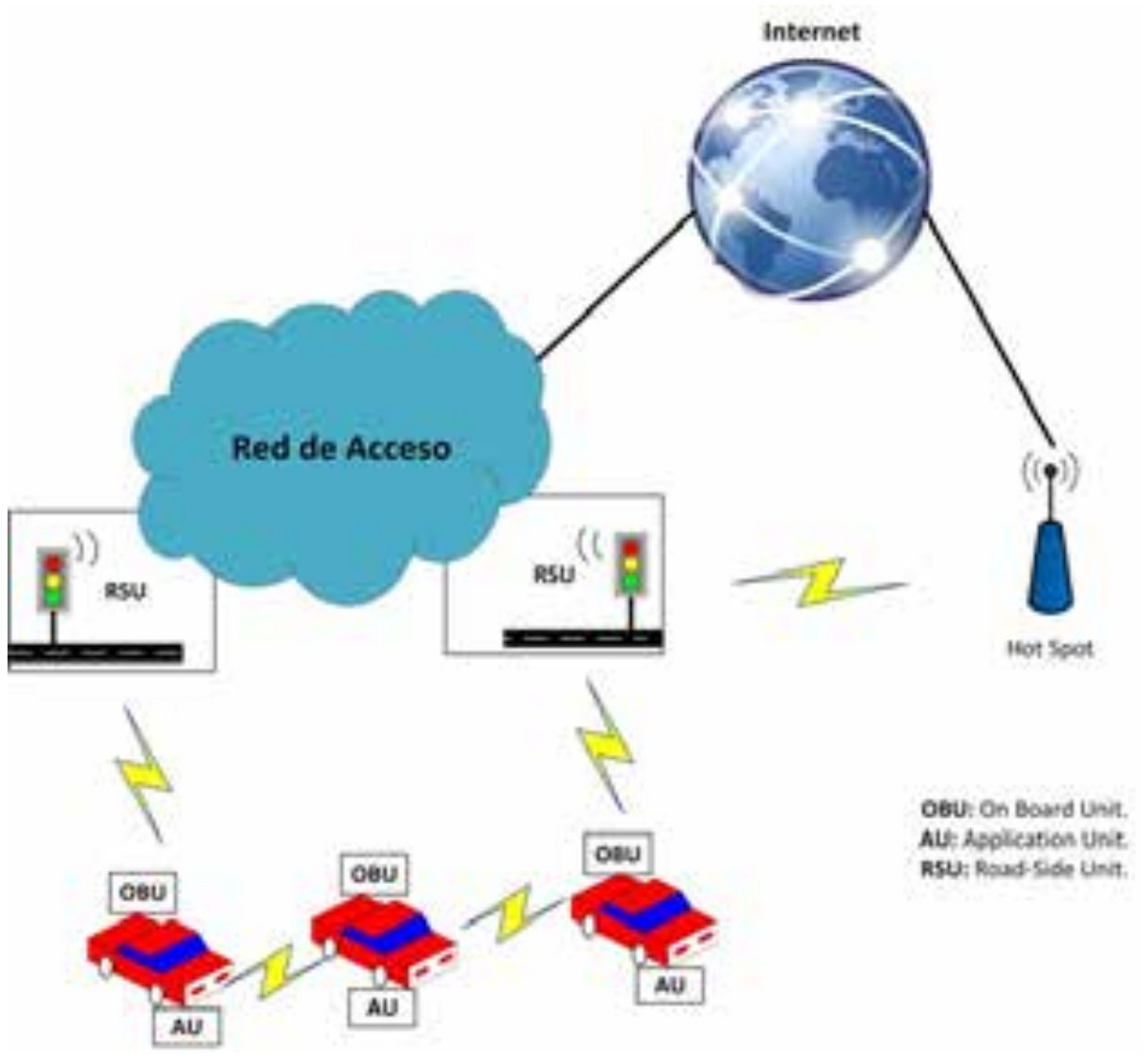

Figura 1: Elementos y dominios de la red vehicular [26]

Otra característica importante de las redes VANETs es la alta autonomía y la auto-gestión que posee este tipo de redes. Ya que cada nodo es autónomo de acceder al medio, transmitir, enrutar y recibir mensajes, sin necesidad de estar enlazado a un centro de control. Es decir, las OBU y RSU gestionan dichas funciones de forma independiente.

Desde el punto de vista del tráfico que manejan estas redes, es importante destacar que las unidades OBU deben tener una capacidad computacional alta para manejar un tráfico de paquetes elevado. Tanto para cursar el tráfico de baja prioridad (por ejemplo, el video streaming) como el tráfico de alta prioridad, pero de baja tasa de datos (como el tráfico de aplicaciones de seguridad). Pero además, las OBU deben enrutar el tráfico que van para otros nodos, ya que también es posible que se establezcan comunicaciones multisalto, en las cuales algunos nodos lejanos se alcanzan a través de sus nodos vecinos.

Las aplicaciones de las VANETs se pueden clasificar en tres tipos: aplicaciones de gestión de la movilidad, de seguridad vial y servicios comerciales y de información. A continuación, se describen cada uno de ellos:

## 3.1 Aplicaciones de gestión de la movilidad

El principal objetivo de estas aplicaciones es mejorar las condiciones de la movilidad mediante el monitoreo y la gestión del tránsito de vehículos. Estas aplicaciones suministran información sobre el flujo vehicular (velocidad, rutas y navegación) y permiten controlar de forma inteligente algunos elementos como los semáforos y los peajes. Además, pueden generar notificaciones a los conductores y las autoridades de transito sobre del estado de las vías, irregularidades e información como placas y licencia de los vehículos.

### 3.2 Aplicaciones en seguridad vial:

Este tipo de aplicaciones tienen como función el monitoreo y recolección permanente de información acerca del estado de las vías para prevenir accidentes. Existen tres subcategorías de este tipo de aplicaciones que son:

- Notificación de señales de tránsito: su función principal es advertir y notificar a los conductores acerca de la señalización vial y brindar cierto tipo de asistencia durante el recorrido.
- Prevención de colisiones: en este tipo de aplicación la RSU detecta el riesgo de colisión que pude haber entre dos vehículos y se informa a los conductores gracias a la OBU.
- Gestión de incidentes: se usan generalmente en situaciones de emergencia ante un accidente de tránsito.

### 3.3 Aplicaciones de información y entretenimiento

Estas aplicaciones principalmente ofrecen información y entretenimiento tanto al conductor y sus pasajeros, mediante el uso de los puntos de acceso WiFi que se encuentran durante el recorrido para: navegar en internet, ver video en streaming, entre otros. Además, permite el intercambio de información sobre sitios de interés como restaurantes, hoteles, estaciones de gasolina así, etc.

## 4. Las antenas impresas o líneas microcinta

Desde 1970, la tecnología microcinta es popular en aplicaciones para microondas y ondas milimétricas. La configuración básica de la tecnología microcinta es similar a la de la tecnología impresa PCB (Printed Circuit Board). La tecnología microcinta (ver figura 2) consta de un sustrato con bajas pérdidas y de una cara inferior y superior que están recubiertas por una capa de cobre [27].

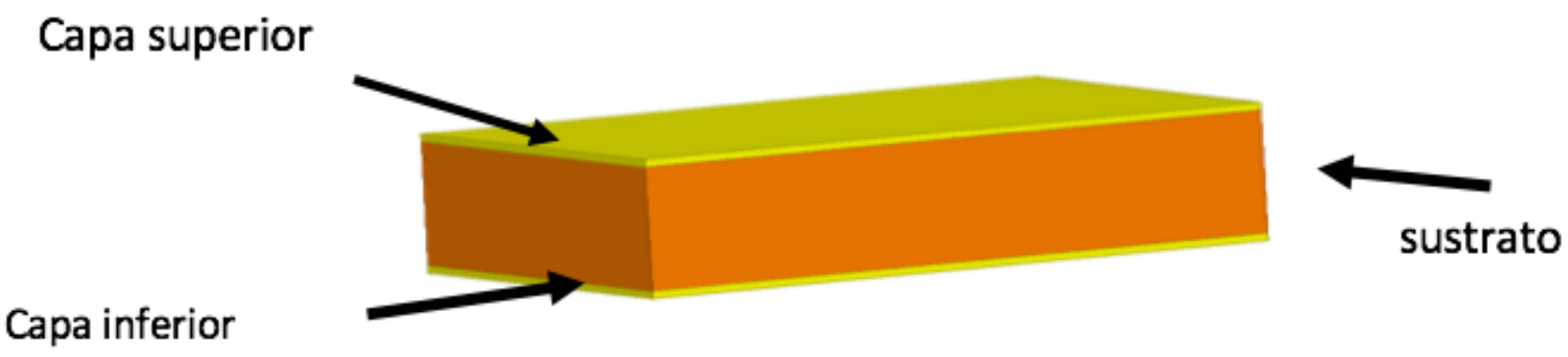

Figura 2: Tecnología microcinta

Entre las antenas fabricadas en tecnología microcinta se encuentran entre otras, las antenas tipo parche y las antenas ranuras. Estas últimas se caracterizan por tener una radiación bidireccional, esto es, puede emitir radiación tanto hacia la parte frontal como hacia la parte trasera [28]. Sin embargo, esta característica puede ser modificada implementando una **cavidad** en su parte frontal o en su parte trasera. Usando ese principio o técnica, en este trabajo se diseñó una **antena ranura** (utilizando tecnología microcinta) en una de las caras metálicas y, además se diseñó una cavidad que fue implementada en el sustrato, utilizando la técnica conocida como guía de onda integrada al sustrato (SIW). A continuación, se hace una breve descripción de la tecnología SIW, con el objetivo de explicar los fundamentos del diseño de una cavidad en esta técnica. Posteriormente, se describirá el diseño completo del circuito que conforma toda la antena, es decir, el diseño de la antena ranura, la cavidad y el acople de estos elementos.

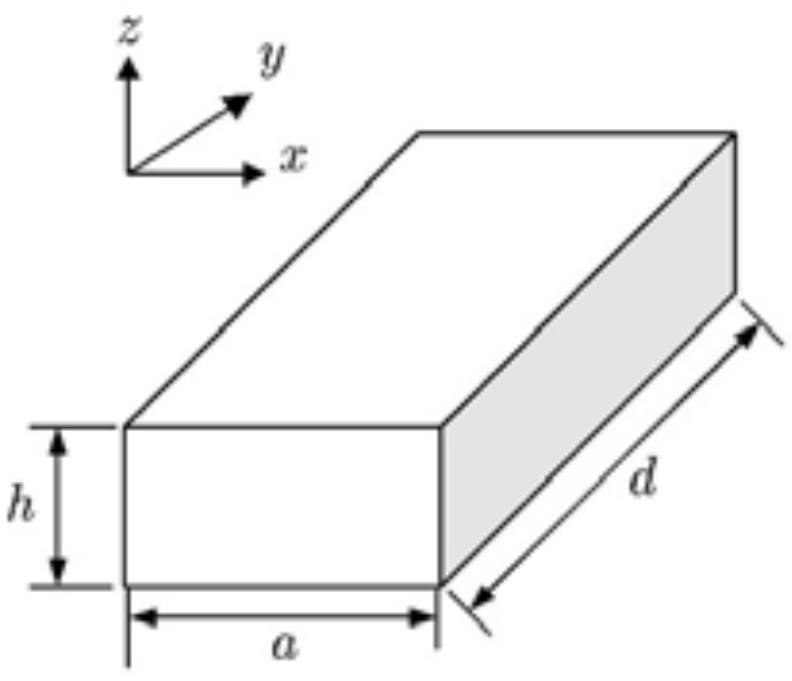

$$a = \frac{c}{2f_c\sqrt{\varepsilon_r}}$$

$$\beta_{mn} = \sqrt{k^2 - \left(\frac{m\pi}{a}\right)^2 - \left(\frac{n\pi}{h}\right)^2}$$

$$k = \omega\sqrt{\mu\varepsilon}$$

$$\lambda_g = \frac{2\pi}{\beta_{mn}}$$

$$d = l\frac{\lambda_g}{2}$$

Figura 3: Diseño de la cavidad [12]

## 4.1 Tecnología SIW

La tecnología SIW (Substrate Integrated Waveguide), consiste en integrar una guía de onda o cavidad (Figura 3) dentro de un sustrato dieléctrico (la placa superior e inferior del sustrato están recubiertas por una capa de cobre) empleando la tecnología impresa de las líneas microcinta (*microstrip line*). La tecnología SIW se diseña a partir de una serie de orificios en forma de cilindros (vías) que atraviesan el sustrato desde la placa inferior hasta la placa superior. Los cilindros están ubicados de tal forma que simulan o son el equivalente de las paredes metálicas de una guía de ondas o una cavidad convencional. El interior de los cilindros debe estar recubierto por una capa de cobre con el fin de obtener la mayor similitud posible a las paredes metálicas de la guía de ondas o cavidad convencional y además para concentrar la energía entre la placa superior e inferior del sustrato. Cada cilindro tendrá un diámetro **d** y una distancia ***dp*** entre cada uno de ellos como se observa en la Figura 4.

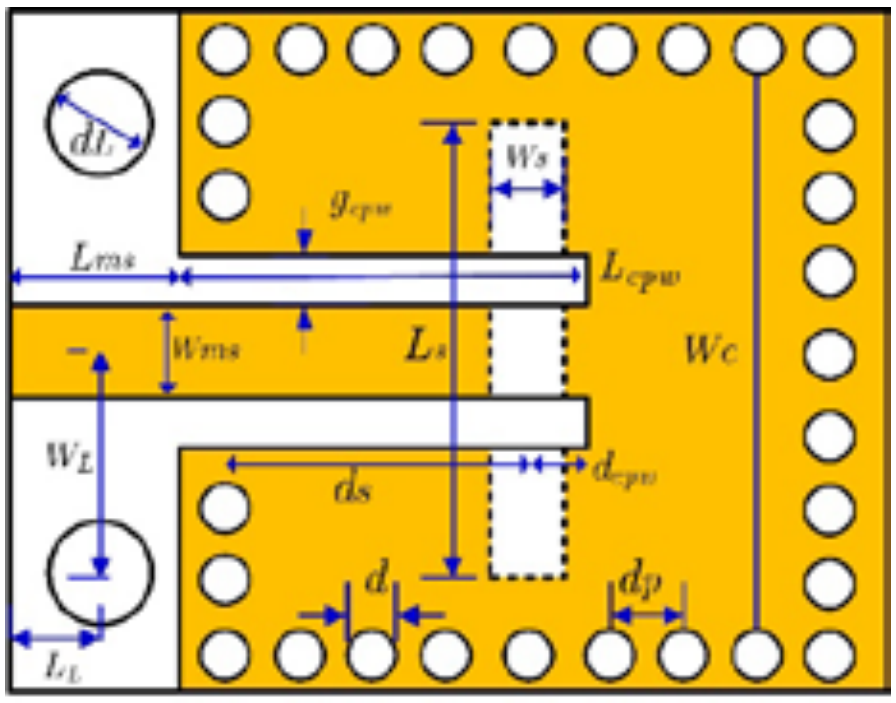

Figura 4: Geometría de las vías [21]

Además de tener dichas dimensiones se debe cumplir las siguientes condiciones que $d/d_p \geq 0.5$ y que $d/\lambda_0 \leq 0.1$, donde $\lambda_0$, es la longitud de onda en el espacio, todas estas condiciones se hacen principalmente para conservar las propiedades básicas de una cavidad general.

Gracias a este tipo de tecnología se pueden reducir costos y materiales, ya que las guías de onda convencionales se fabrican con piezas de metal grandes y pesadas que con la tecnología SIW se omiten. Sin embargo, se siguen conservando las propiedades de propagación y pérdidas de dicha cavidad. Otra ventaja de la tecnología SIW es que la construcción se hace más sencilla ya que se construye a partir de técnicas de circuitos impresos.

Las antenas tipo SIW, al ser guías de onda, necesitan una alimentación que, por sus propiedades, puede ser de la forma convencional, por ejemplo, una línea microstrip. Este tipo de alimentación emplea la misma técnica de fabricación que las SIW, es decir, la tecnología impresa PCB. Para garantizar una correcta alimentación se debe hacer mediante un acople entre una guía de onda coplanar (CPW, Coplanar Waveguide) y una línea microstrip. Este acople funciona de manera adecuada debido a que las líneas de campo eléctrico de ambas estructuras tienen la misma orientación [29].

# 5. Diseño y simulación de la antena integrada en sustrato

## 5.1 Diseño del circuito

En esta sección se presenta el diseño completo de una nueva antena SIW. El circuito que conforma la antena SIW incluye los siguientes componentes: una antena ranura, la línea de alimentación de la cavidad (línea microstrip) y la cavidad (también conocida como CPW). Uno de los objetivos de diseño de la cavidad es aprovechar las propiedades del modo $TE_{101}$ para obtener la cavidad más pequeña posible.

En primer lugar, el diseño de la cavidad se realiza con las ecuaciones de la Figura 4. El método comienza por la definición de la frecuencia de resonancia $(f_0)$ de una cavidad que resuena en su primer modo $TE_{101}$. En este trabajo se estableció una frecuencia de resonancia $f_0 = 2400MHz$. Desde el punto de vista físico, la cavidad diseñada puede ser vista como una guía de onda rectangular cortocircuitada en ambos extremos. En la Figura 4 se muestra una representación simplificada de dicha cavidad. Como condición importante se debe tomar la frecuencia de corte $(f_c)$ cuando la cavidad opera en el modo fundamental $TE_{10}$. El objetivo es diseñar la cavidad con la frecuencia de resonancia en el modo más alto y tan lejos como sea posible de $(f_0)$. A continuación, el procedimiento es el siguiente: el ancho *(a)* de la cavidad se calcula con la frecuencia $(f_c)$ y la altura del sustrato *(h),* la cual define la altura de la guía de onda. A partir de estos valores, se puede determinar la constante de propagación $(\beta_{mn})$ y la longitud de onda guiada $(\lambda g)$, así como la longitud *(d)* de la cavidad para resonar en el primer modo $TE_{101}$ [30]. En la Figura 5 se ilustra la diferencia entre el modo fundamental $TE_{101}$ y los modos de orden superior en función de la diferencia $f_0 - f_c$. Para trazar esta curva, el valor de $f_c$ se varió mientras que el valor de $f_0$ permaneció fijo. De la Figura 5 se puede determinar que los modos superiores más próximos al $TE_{101}$ son el $TE_{102}$ y el $TE_{201}$. La cavidad fue diseñada utilizando el sustrato Duroid 5880 $(\varepsilon_r = 2.2, tan\delta = 0.0009, h = 1.575\ mm)$ y se realizó una cavidad cuadrada de 59,550818 mm con $f_0 - f_c = 700\ MHz$. Las paredes laterales de la cavidad se llevaron a cabo usando cilindros metalizados llamados *vías* sobre el sustrato. El espacio entre cada cilindro y diámetro se seleccionaron de acuerdo a la metodología planteada en [31], con el fin de conservar las propiedades de la cavidad empleado tecnología SIW.

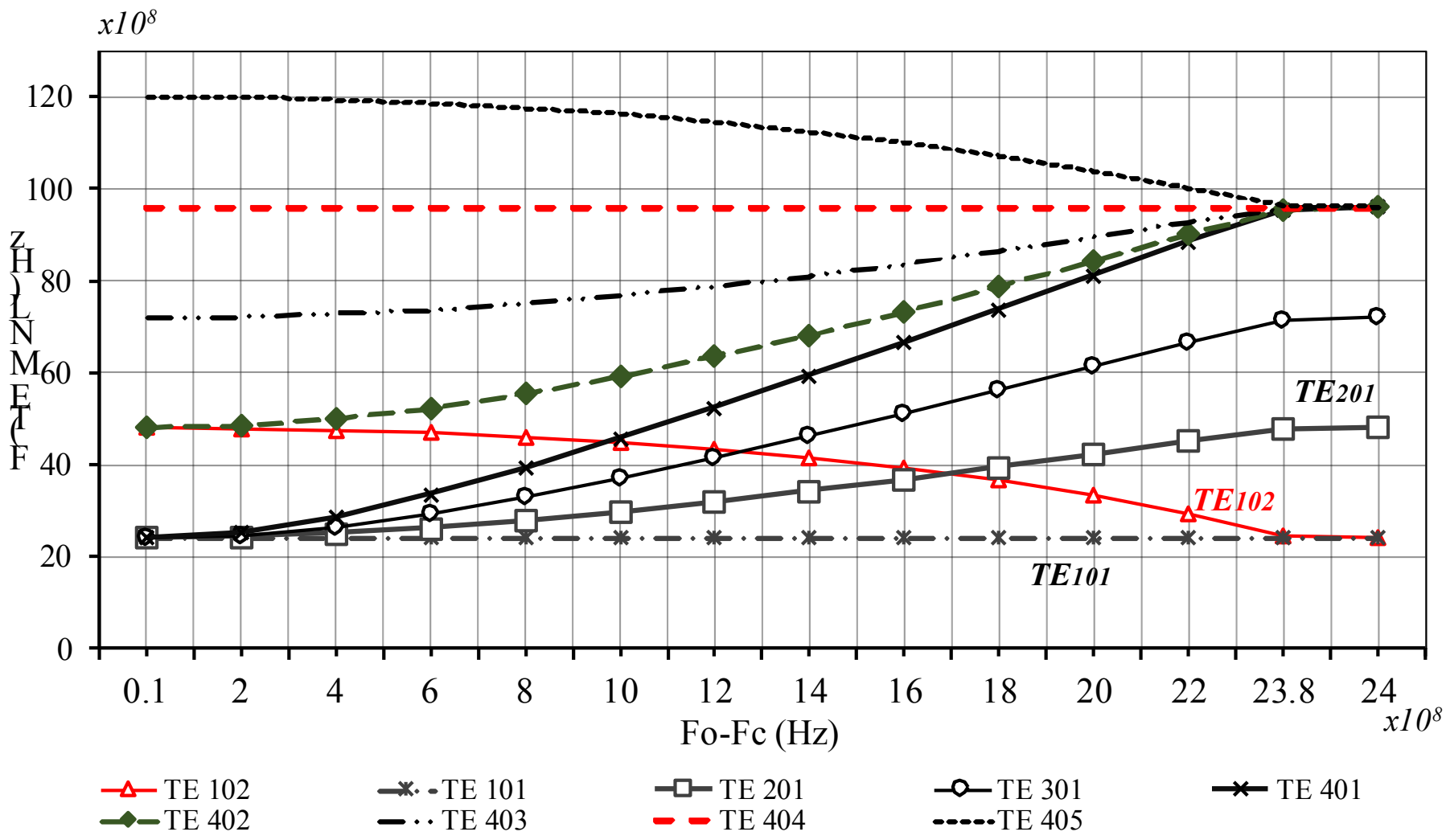

Figura 5: Proximidad de los modos

## 5.2 Acoplamiento a la cavidad usando el acceso CPW

Al realizar el acoplamiento se obtiene una estructura como la mostrada en la Figura 6, donde se puede apreciar que la línea microstrip y la CPW, deben estar perfectamente acopladas. En la Figura 6 se muestra una vista en tres dimensiones de la composición del circuito diseñado y la Figura 7 describe las dimensiones que deben ser ajustadas para diseñar el circuito. Como se observa en la Figura 7, en la cara superior de la cavidad se utiliza una transición de microcinta a la cavidad de la SIW. Esta transición consiste de una línea microstrip de 50 Ω y de longitud *Sel* que se acopla a una guía de ondas coplanar (CPW) de acceso a la cavidad.

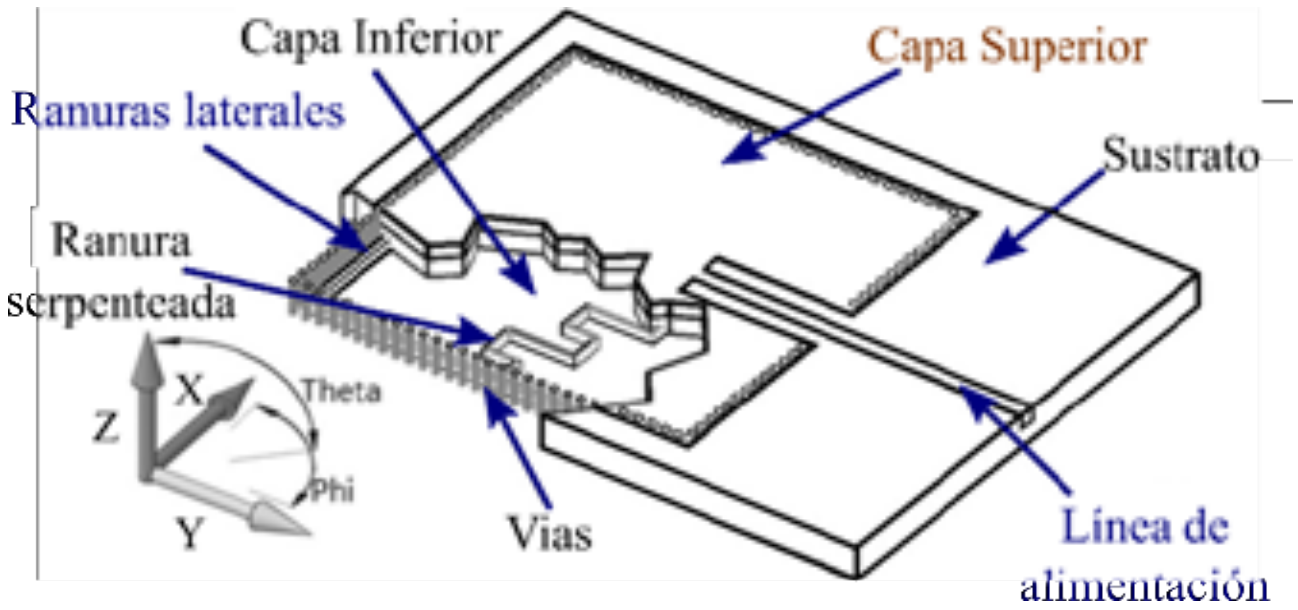

Figura 6: Vista 3D del circuito implementado [12]

La transición de microcinta a CPW, puede ser diseñada debido a que el campo eléctrico de la microcinta y la CPW tienen la misma orientación. La longitud *Cl* es el parámetro que se utiliza para convertir el modo cuasi-TEM de la línea microcinta al modo $TE_{101}$ de la cavidad de la SIW. A medida que los campos eléctricos de las dos estructuras tienen la misma orientación y perfil, el diseño de transición es fácil de obtener. Además, el factor de calidad externo depende principalmente de la distancia *Cl* y la anchura *CW*.

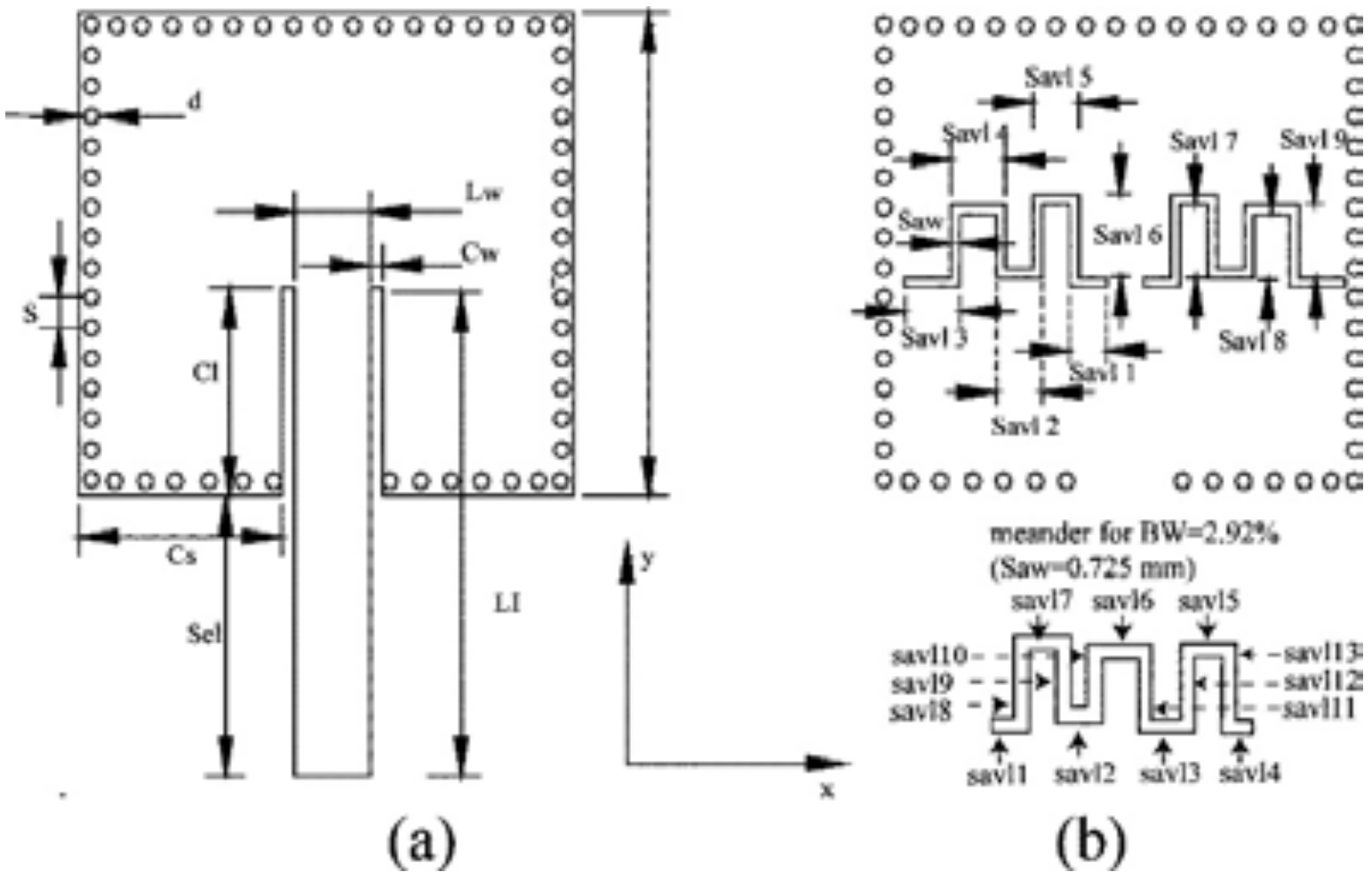

Figura 7: Acoplamiento microstrip y CPW (a) Vista desde arriba (b) Vista del plano inferior [12].

## 5.3 Diseño de la Antena

Tal como se mencionó en la seccione previas, la antena diseñada corresponde a una antena serpenteada, comúnmente llamada antena de ranura (slot antenna). Este tipo de antenas son implementadas sobre la placa metálica que está en la parte inferior de la cavidad (ver Figura 6 y Figura 7) y consiste de una ranura de forma serpenteada. Para obtener el principio de radiación en la antena de ranura serpenteada, es útil tener en cuenta las líneas de corriente de la cavidad alrededor de la ranura. Estas, líneas de corriente pueden ser descompuestas en dos partes. La primera parte circula alrededor de la ranura y es la principal responsable de la creación de una condición de resonancia. La segunda es principalmente responsable de la radiación y fluye perpendicular a la ranura. La Figura 8 muestra los componentes anteriormente mencionados.

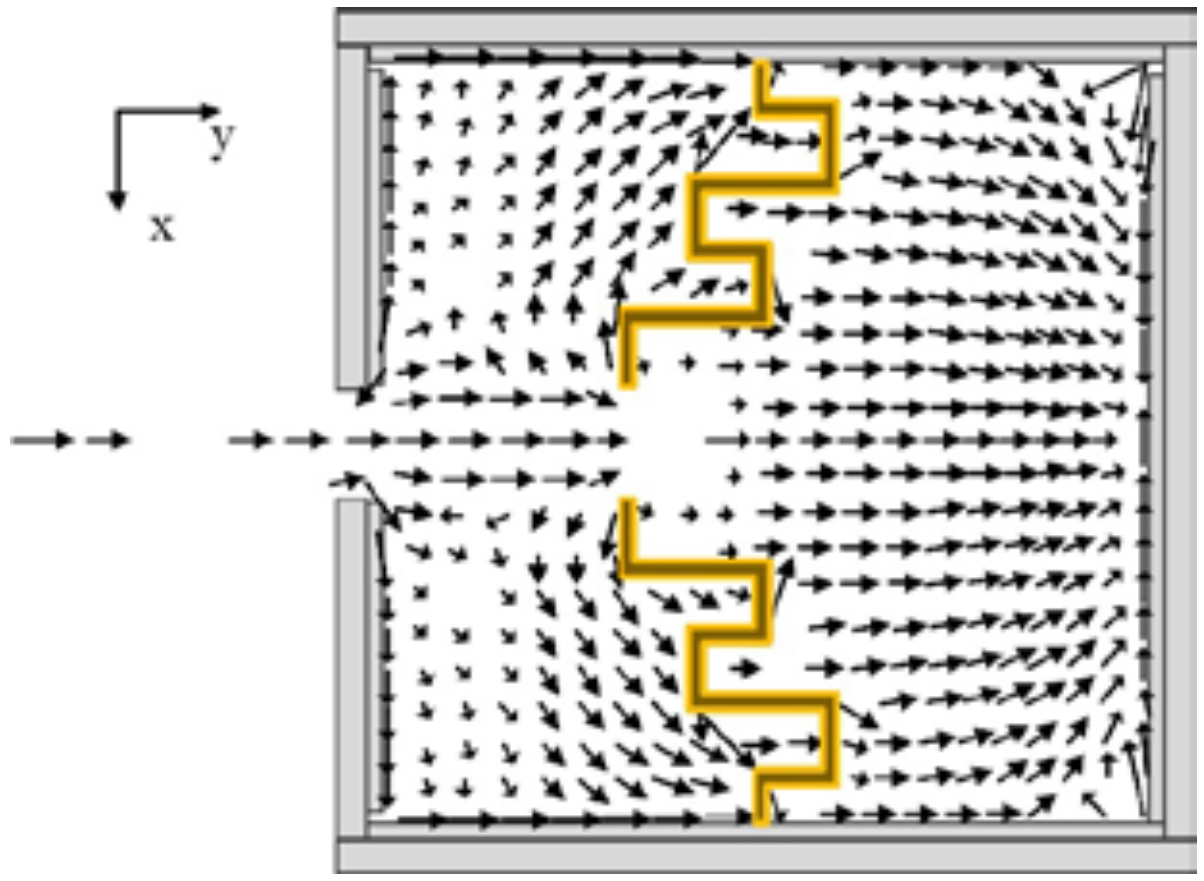

Figura 8: Distribución de corrientes en la antena serpenteada [12].

La ranura serpenteada se diseñó con una longitud total de $\lambda/2$. Esta longitud es establecida para que la ranura pueda adaptarse a la anchura de la cavidad, evitando que sobrepase y se salga del marco creado. La ranura serpenteada cuenta con ranuras laterales, las cuales son usadas para cargar inductivamente la antena. Además, tienen propiedades como permitir reducir la frecuencia de resonancia y cambiar la resistencia de radiación.

Las longitudes de las secciones más largas y más cortas de la ranura serpenteada se calcularon experimentalmente y fueron fijadas en $\lambda/16$ y $\lambda/32$, respectivamente. El ancho de cada línea de la ranura serpenteada está dado por 1/10 de su longitud.

## 5.4 Resultados de la simulación de la antena

En la Figura 9 está representado el parámetro de reflexión (S11) de la antena. Se puede observar que son obtenidas una frecuencia de resonancia de 2.398 GHz y un ancho de banda de 20 MHz, siendo estos dos valores óptimos para ser utilizados en la banda de frecuencias propias de los estándares WI-FI. Las simulaciones electromagnéticas se hicieron en el software Ansoft High-Frequency Structure Simulator (HFSS). Todas las dimensiones de la antena se describen en la Tabla **1**, las cuales siguen la estructura de la Figura 7.

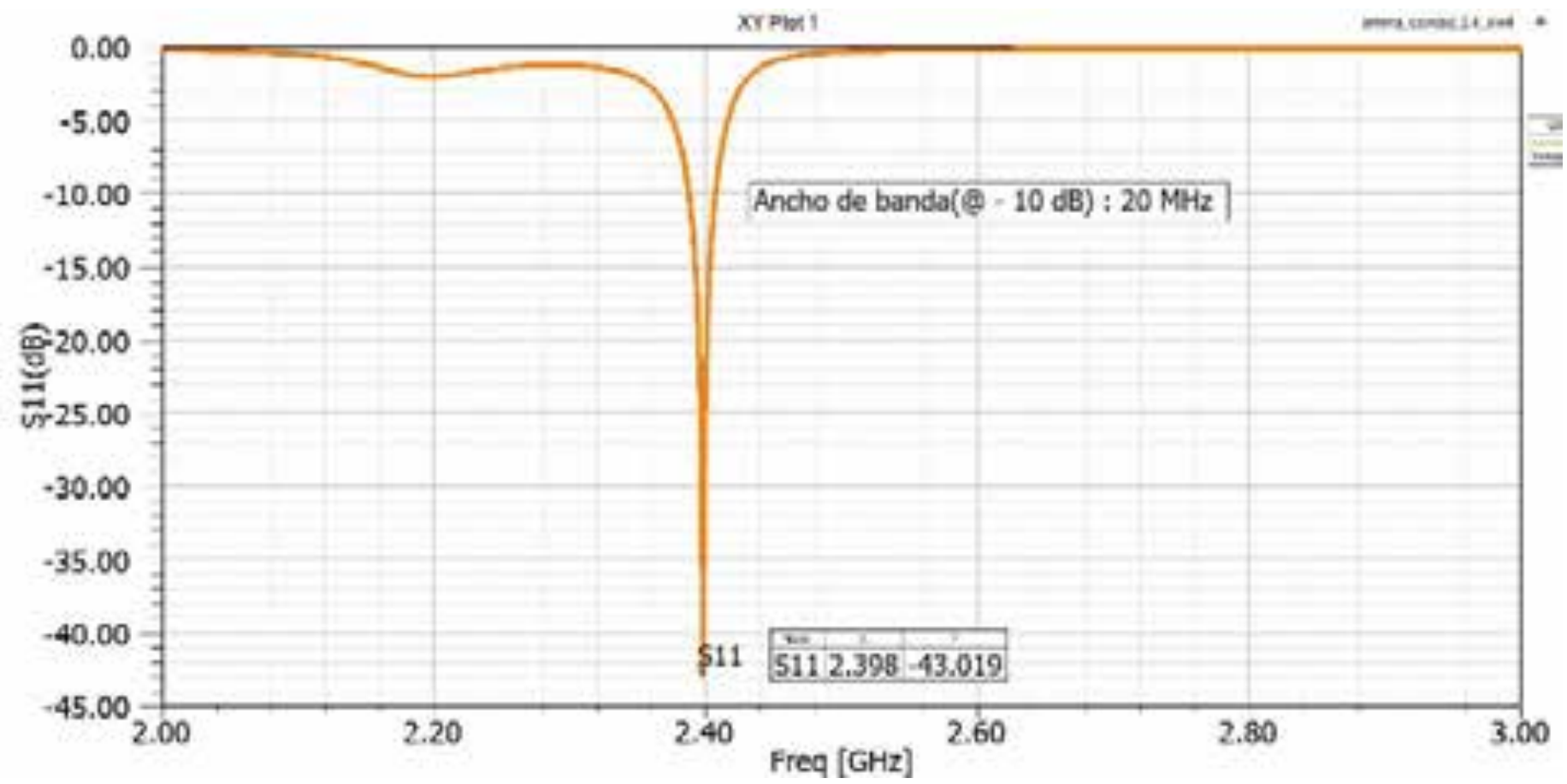

Figura 9: Frecuencia de resonancia

**Tabla 1 Dimensiones de la antena**

| MEDIDA | VALOR (mm) | MEDIDA | VALOR (mm) |
|---|---|---|---|
| **W** | 59.550818 | **S** | 1.972 |
| **LW** | 4.9 | **Saw** | 1.25 |
| **L1** | 44.609 | **Sav11** | 4.95 |
| **CW** | 0.4 | **Cs** | 28.025 |
| **SAV17** | 12 | **Sav18** | 10.76 |
| **C1** | 29.61 | **Sel** | 15 |
| **SAV12** | 6.2 | **Sav14** | 7.4 |
| **SAV13** | 7.46 | **Sav15** | 6.21 |
| **D** | 1 | **Sav16** | 11.161 |
| | | **Sav19** | 9.92 |

En la Figura 10 se observa el modelo de la antena que fue simulado. Además de la frecuencia de operación, también mediante simulación se obtuvo el patrón de radiación (Figura 11). El patrón de radiación es obtenido en el plano de la antena ranura serpenteada. La ganancia de la antena fue de aproximadamente 4dB. Como se observa en la Figura 11 toda la radiación de la antena es radiada hacia la parte inferior de la antena, esto es debido a que la cavidad que se forma a partir de la tecnología SIW impide que se irradie hacia la parte superior de la antena.

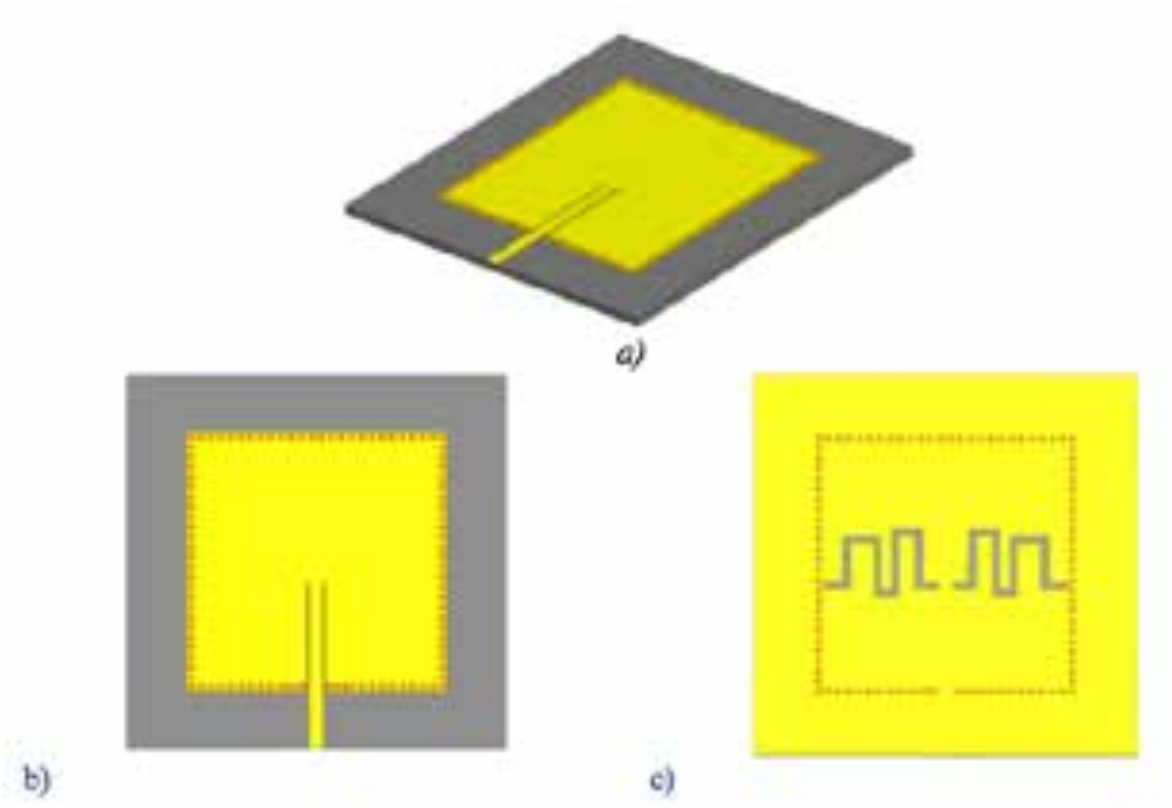

Figura 10: Antena diseñada y simulada en HFSS, a) vista 3D, b) plano superior, c) plano inferior

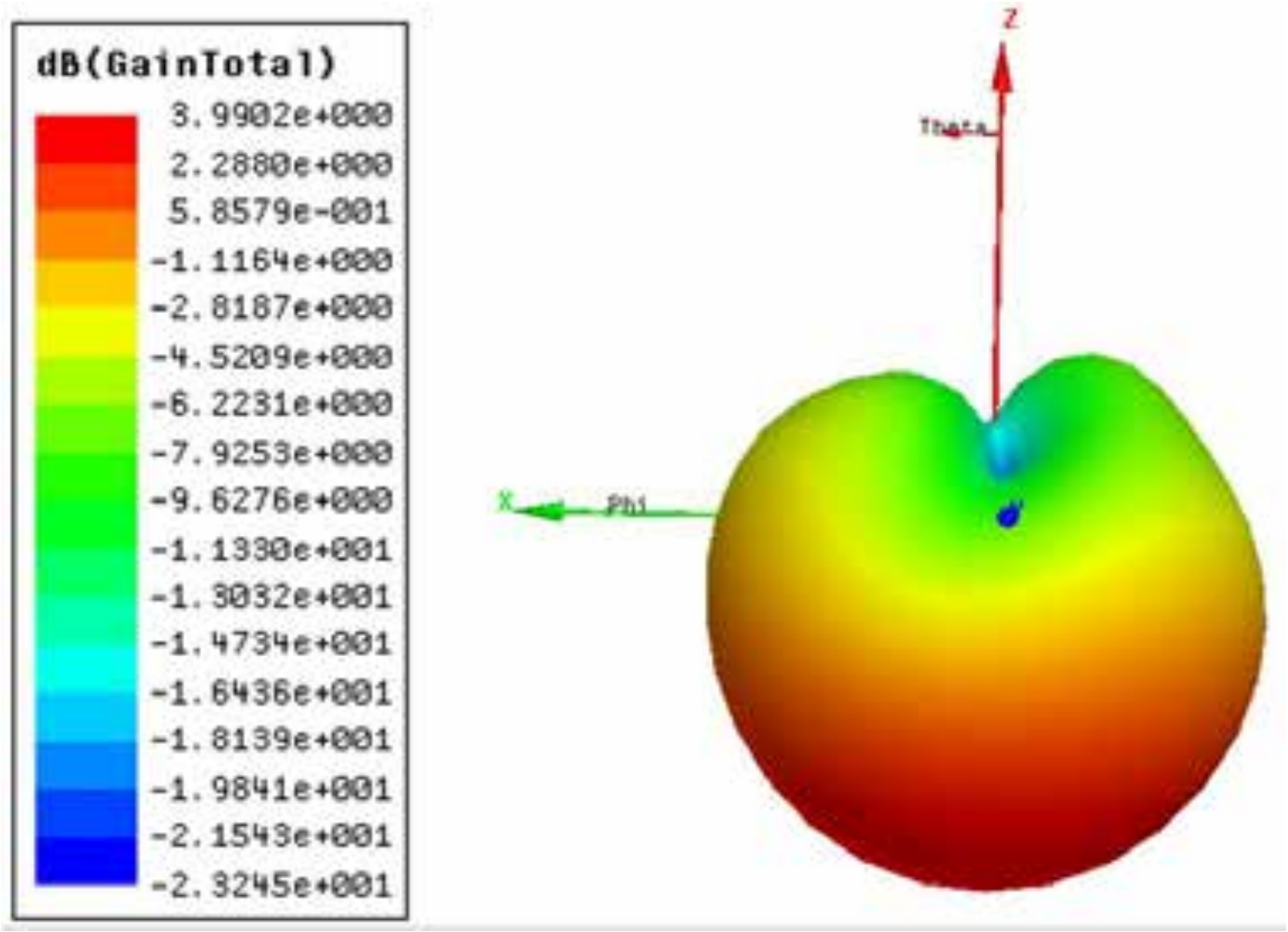

Figura 11: Patrón de radiación de la antena simulada.

## 6. Simulación de redes vehiculares ad hoc

Una vez realizado el diseño y la simulación de la antena SIW, se hace una descripción de las herramientas y la metodología para analizar el funcionamiento de una VANET, a través de simulaciones. Es a través de estudios basados en simulaciones, que se evaluó la integración de la antena SIW con los vehículos que conforman una VANET. En la **¡Error! No se encuentra el origen de la referencia.**, se describe de forma gráfica los elementos utilizados para la configurar las simulaciones.

Uno de los ambientes más utilizados para la evaluación de las VANETs son las plataformas de simulación, debido a los altos costos que supone la experimentación en un sistema real. Entre los simuladores más extendidos por la comunidad científica está el simulador NS2. Sin embargo, este software es un simulador de redes de comunicaciones que carece de algunos elementos importantes para la configuración y la evaluación de las VANETs. Por ejemplo, las rutas a través de las cuales se van a desplazar los diferentes vehículos no pueden ser creadas por NS2, ya que este software no tiene las herramientas necesarias para configurar dichas rutas con el diseño y la geometría de un sistema típico de carreteras propio de una ciudad. Para la generación de estas rutas se utilizó una herramienta especializada llamada MOVE, la cual es una aplicación JAVA que permite generar las rutas de desplazamiento, las calles, semáforos, y la cantidad de automóviles que se quiere simular.

Además de las rutas, es necesario crear los modelos o patrones de tráfico, es decir, todos los movimientos que realizan los vehículos de una VANET. Para el estudio de simulación que se presenta en este artículo, los modelos de tráfico fueron generados con la aplicación SUMO. Estos modelos incluyen el desplazamiento de los vehículos, los giros realizados en los cruces de vías, los adelantamientos, las paradas en semáforos y todos los demás aspectos relacionados con la movilidad de los vehículos.

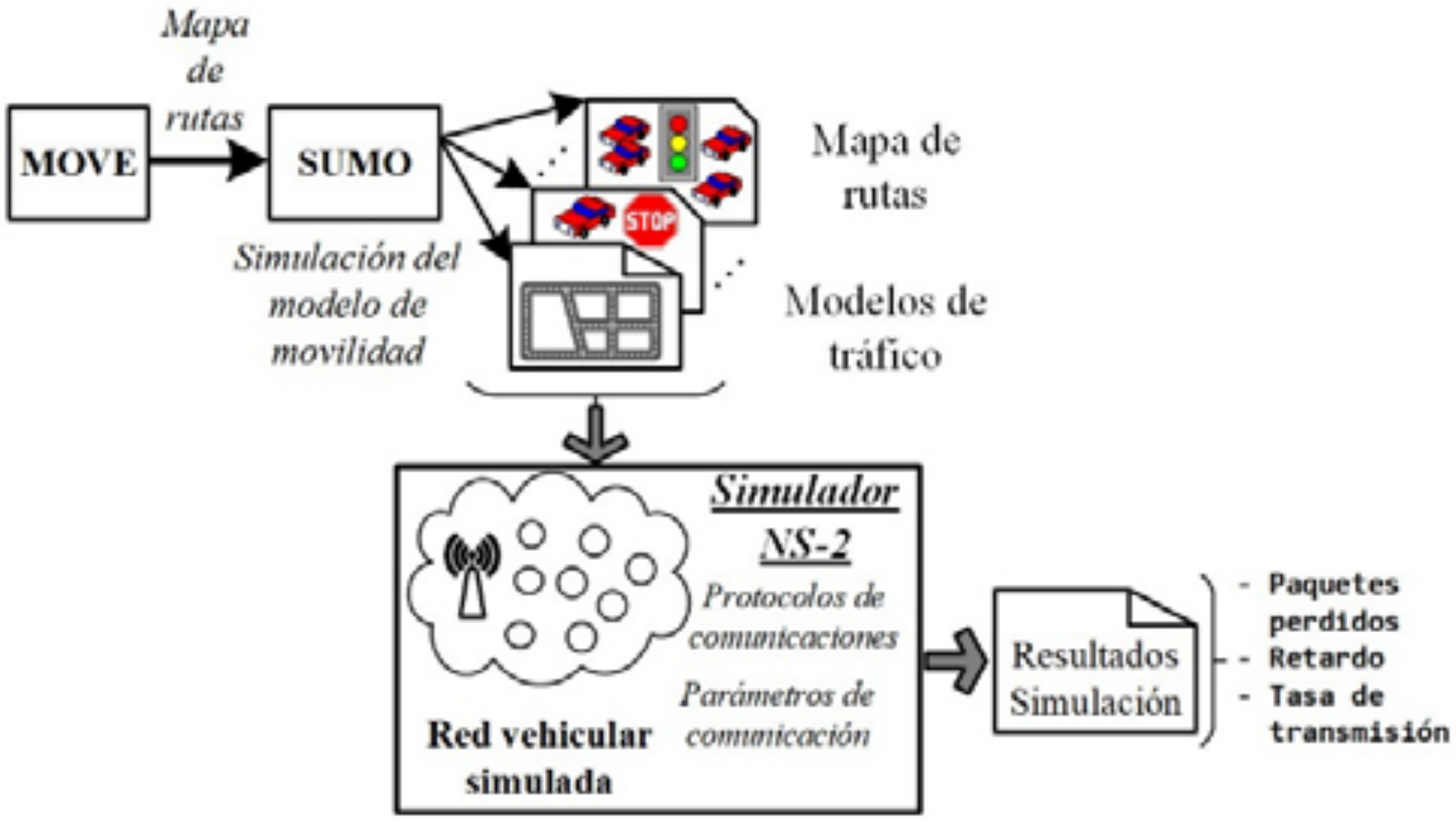

Figura 12: Esquema general de la metodología utilizada para la configuración de las simulaciones.

## 6.1 Escenario de simulación: visualización del mapa de rutas y el tráfico de los vehículos

Las rutas generadas en MOVE y los patrones de tráfico pueden ser visualizados en la interfaz gráfica de SUMO. En la Figura 10 se muestra una captura de pantalla de SUMO, donde se visualiza el mapa de rutas generado (Figura 13a) y también una captura de pantalla con acercamiento para visualizar algunos de los vehículos que conforman el tráfico que circula por el mapa (Figura 13b).

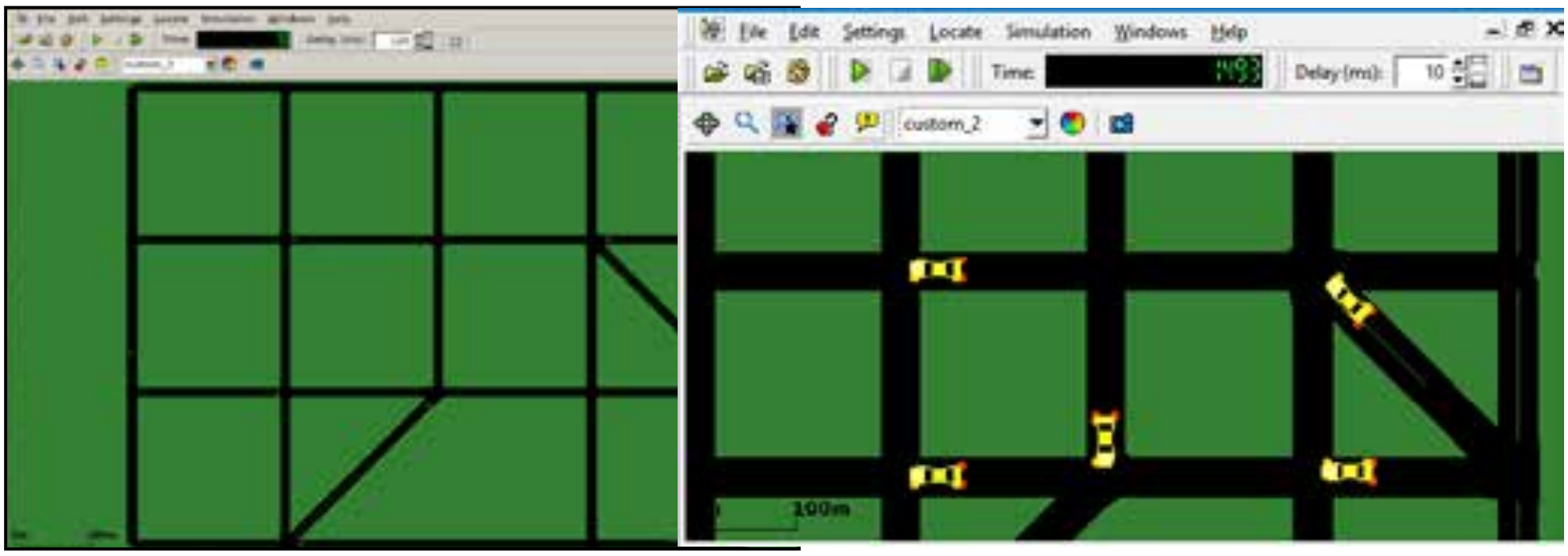

*(a)* *(b)*

Figura 12: (a) Mapa generado con MOVE y SUMO, (b) Acercamiento para visualizar algunos vehículos circulando por las calles

Por medio de la interfaz gráfica de SUMO, mostrada en la Figura 13a, también se puede realizar la simulación de la movilidad de los vehículos por el mapa durante determinado tiempo. Esto genera varios archivos que posteriormente son integrados en NS2 para realizar la simulación completa de la VANET, es decir, integrando las rutas, la movilidad y las comunicaciones entre los vehículos.

## 6.2 Evaluación de la VANET en NS2

Con las rutas y los modelos de tráfico generados con MOVE y SUMO, el siguiente paso fue integrar estos datos a la simulación de la VANET en NS2. También se configuraron las antenas de los vehículos con los parámetros de la antena integrada al sustrato, tales la potencia de transmisión, tipo de antena, frecuencia de operación, ancho de banda y ganancia.
Además de estos parámetros, se configuraron en NS2 los parámetros necesarios para el establecimiento de los enlaces inalámbricos y la transmisión de los datos dentro de la VANET. Algunos de los parámetros utilizados se muestran en la Tabla 1.

En la Figura 13 se muestra una captura de pantalla de NS2 durante la ejecución de la simulación de la VANET. En dicha captura de pantalla se visualizan, tanto los diferentes nodos que componen la VANET, como el intercambio de mensajes entre ellos. Este tráfico en la VANET se configuró como tipo CBR (Constant Bit Rate) sobre UDP con un tamaño de paquete de 512 bytes.  La fuente y el destino del tráfico fue seleccionado aleatoriamente y transmiten a una tasa de 500Kbps.

**Tabla 2 Algunos parámetros de simulación**

| PARÁMETRO | VALOR |
|---|---|
| **Frecuencia de operación** | 2.4 GHz |
| **Potencia** | 0.28 Watts |
| **Protocolo de capa MAC** | IEEE 802.11 |
| **Capacidad de transmisión de los enlaces** | 11Mbps |
| **Protocolo de enrutamiento** | AODV |
| **Número de nodos** | 15 |
| **Área del escenario** | 820m x 620m |

Las métricas evaluadas en la simulación fueron: el flujo de datos (o throughput) y el porcentaje de paquetes perdidos. Estas métricas permiten identificar, cada instante de tiempo, el estado de los enlaces de comunicaciones y la calidad de los mismos.

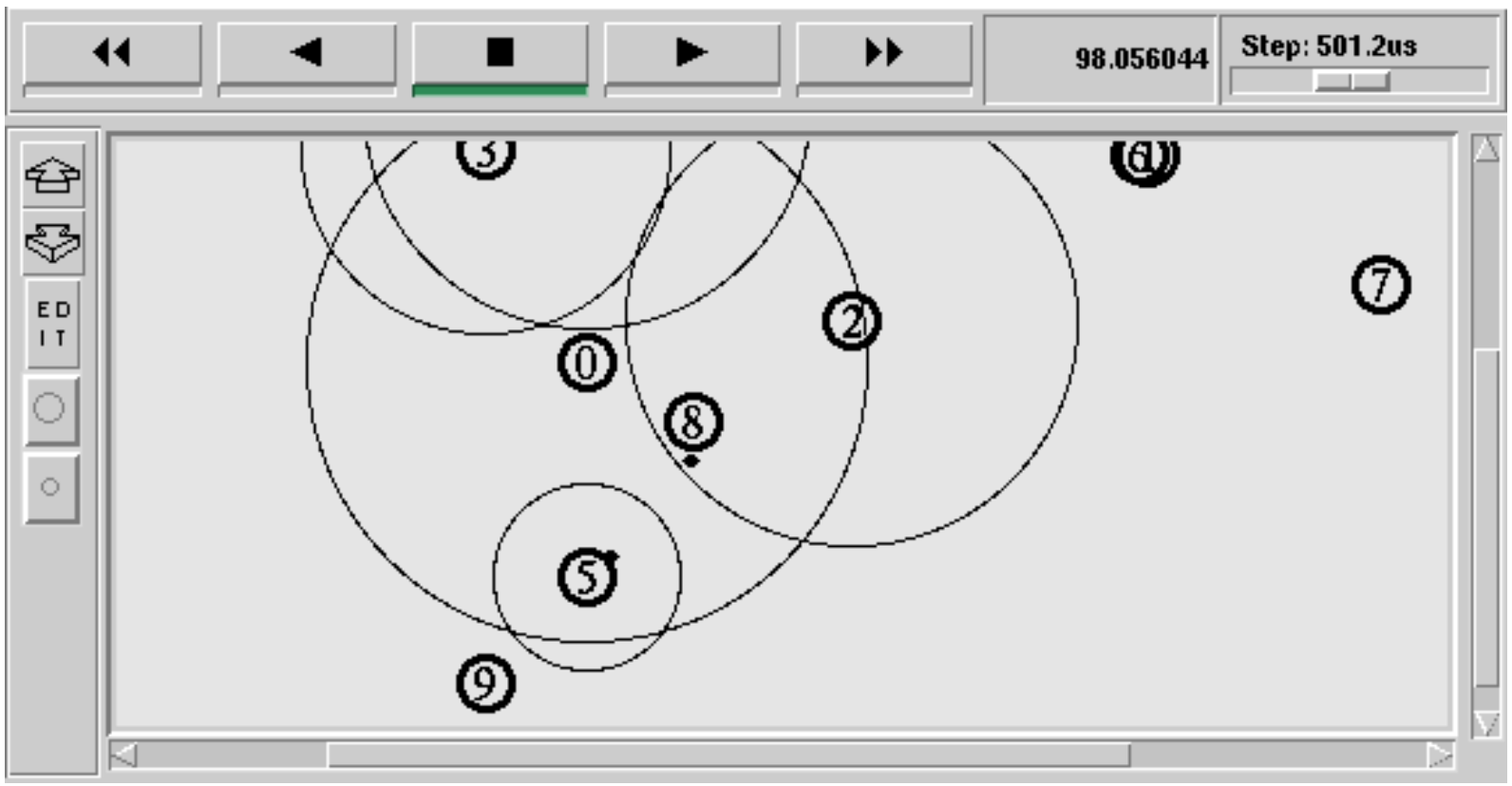

Figura 13: Visualización parcial del escenario de la VANET en NS2

### 6.3 Resultados de la simulación de la VANET

A continuación, se presentan los resultados obtenidos en NS2, después de simular la red VANET, cuyos nodos tienen configurada la antena SIW diseñada. En primer lugar, el flujo de datos (throughtput) obtenido durante el tiempo de simulación (ver Figura 14) muestra que se generaron dos pérdidas de conectividad: una en el instante 105s y otra en t=120s. Esto se debió a que, en dichos instantes, los nodos que intervenían en la transmisión de los datos de alejaron y perdieron la conectividad entre sí. Sin embargo, el enlace fue reestablecido, a través de otros nodos de la red, conformando un enlace multisalto. El tiempo que tardó la red en restablecer cada desconexión fue de 0.7s y 2s respectivamente. Estos tiempos de reconexión pueden ser significativos si se está transmitiendo información sensible al retardo como un flujo de video. Los tiempos de reconexión dependen principalmente del protocolo de enrutamiento utilizado, lo cual debería analizarse con más profundidad en otros estudios. Adicionalmente, en la Figura 14 se observa que se presentaron dos instantes en los cuales se generaron unas subidas instantáneas del throughput. Estas subidas ocurrieron cuando se establecieron las rutas por primera vez (instante de tiempo 30s) y cuando se produjo la reconexión después de una pérdida de conectividad (t=120s). Este incremento del flujo de datos se debe al aumento de los paquetes de señalización que genera el protocolo de enrutamiento AODV, con el fin de encontrar nuevas rutas.

En la Figura 15 se observa que durante los intervalos en que se registraron las pérdidas de conectividad, también se presentó una alta pérdida de paquetes. Adicionalmente, se registró una alta pérdida de paquetes al inicio de la transmisión de datos (instante 30s). Estas pérdidas se deben a una alta latencia en el establecimiento de las rutas por parte del protocolo de enrutamiento. Es importante anotar que esta latencia podría ser disminuida si, en lugar de utilizar un protocolo reactivo (bajo demanda) como AODV, se utiliza un protocolo proactivo como OLSR.

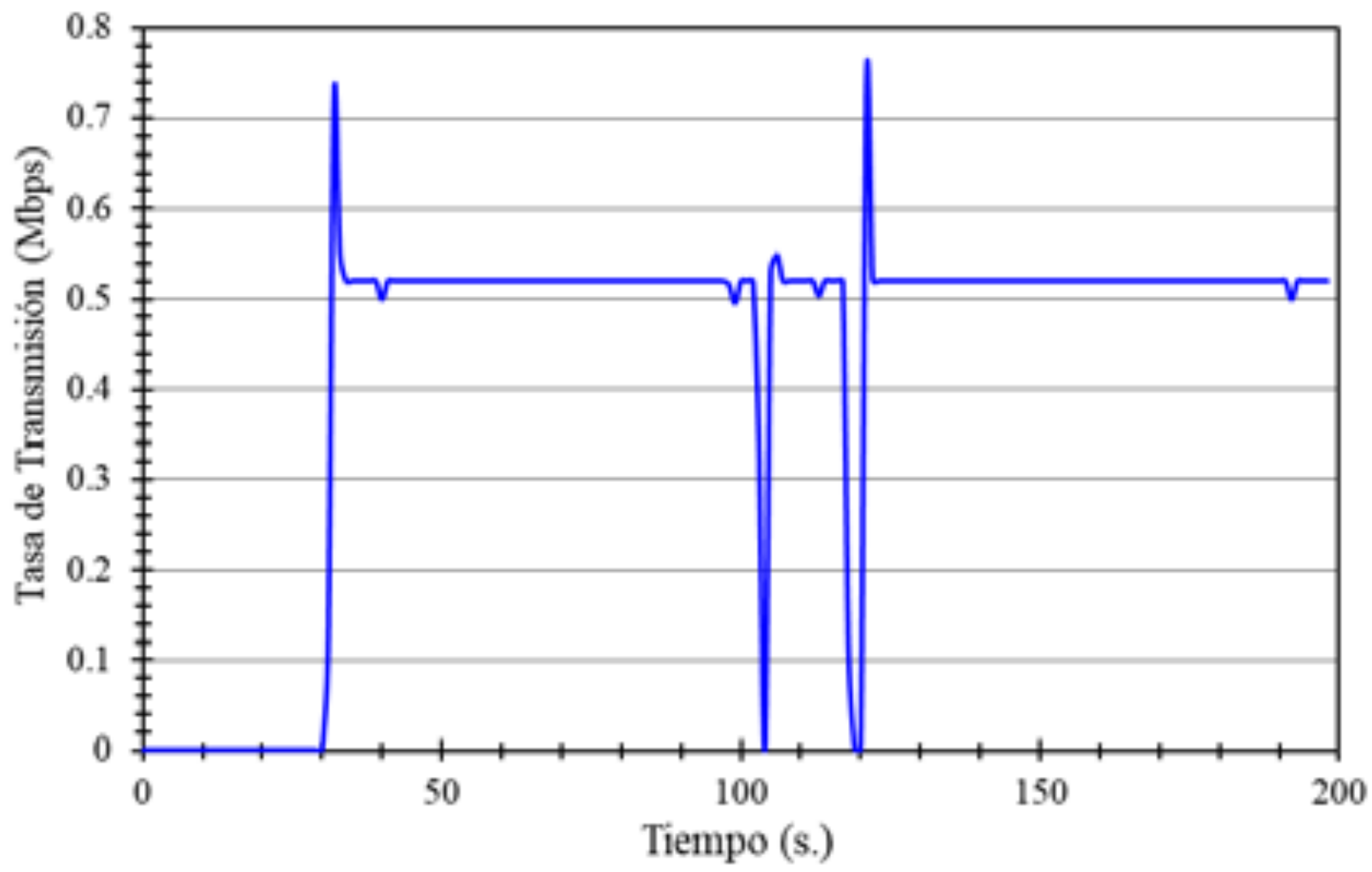

Figura 14: Flujo de datos (Throughput)

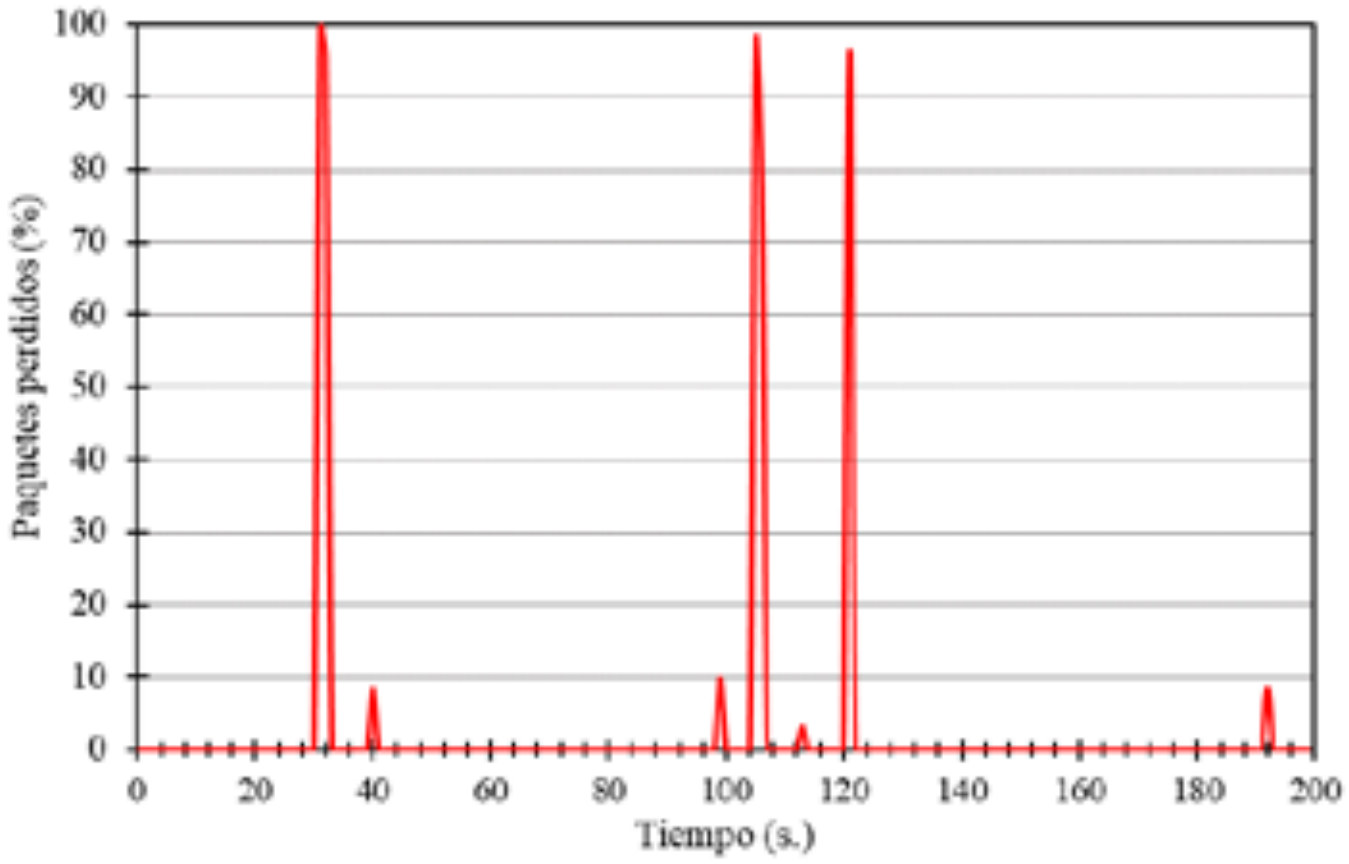

Figura 15: Paquetes perdidos

Mientras que el enlace se mantuvo activo entre los nodos de la VANET, las pérdidas de paquetes estuvieron por debajo del 10%, esto muestra que los parámetros establecidos para la comunicación, entre ellos los parámetros de la antena SIW, funcionaron correctamente. Esto a pesar de la movilidad de los nodos y del establecimiento de rutas multisalto entre el origen y el destino, que disminuyen significativamente el máximo throughput que puede ser alcanzado y aumentan el riesgo de congestión en los nodos, tal como se evidencia es algunos estudios realizados previamente [32].

## 7. Conclusiones

Uno de los aspectos claves para la implementación efectiva de una VANET es la construcción de los nodos inalámbricos con los requerimientos electrónicos adecuados, en términos de potencia, eficiencia energética, etc. En particular, el diseño de las antenas de los nodos inalámbricos es un aspecto que ha despertado creciente interés. Con relación a este aspecto, en este trabajo se propuso una antena integrada al sustrato que puede ser integrada a los nodos que conforman una VANET. La antena propuesta fue evaluada por medio de simulación y los resultados muestran que esta resuena a la frecuencia a la cual se diseñó inicialmente la cavidad. Además de que se obtuvo un ancho de banda amplio que es suficiente para la transmisión de datos de las VANET.

Después de diseñar y evaluar la antena, se realizó la simulación de una VANET, cuyos nodos tenían configurados los parámetros de la antena SIW diseñada. El estudio de simulación fue desarrollado con el propósito de evaluar la integración de la antena SIW en los nodos de una VANET y determinar la conveniencia de su uso en este tipo de redes. Para este fin se evaluaron algunos parámetros de calidad de servicio (pérdidas de paquetes y throughput). Los resultados de las simulaciones muestran que el flujo de datos y las pérdidas obtenidas no son atribuibles a la antena SIW, sino por el contrario, se demostró que la antena permitió el establecimiento de conexiones entre los nodos inalámbricos a una tasa y ancho de banda adecuados para la transmisión de los datos. También se pudo determinar la influencia que ejerce la movilidad de los nodos en las pérdidas de paquetes y en la cantidad de datos que pueden ser transmitidos. Este hecho impone numerosos retos en términos de enrutamiento y en mecanismos de calidad de calidad de servicio, sobre todo cuando se transmite tráfico sensible a las variaciones del retardo y las pérdidas.

## 8. Referencias